\documentstyle[twocolumn,psfig,floats,aps]{revtex}

\begin{document}
\twocolumn[\hsize\textwidth\columnwidth\hsize\csname
@twocolumnfalse\endcsname

\author{Andreas~L\"auchli$^1$, Didier~Poilblanc$^{1,2}$, 
T.~M. Rice$^1$ and Steven~R.~White$^3$}
\title{Li Induced Spin and Charge Excitations in a Spin Ladder}
\address{$^{1}$Institut f\"ur Theoretische Physik, ETH-H\"onggerberg,
  CH--8093 Z\"urich, Switzerland.\\
$^2$Laboratoire de Physique Quantique \& UMR--CNRS 5626,
Universit\'e Paul Sabatier, F-31062 Toulouse, France \\
$^3$Department of Physics and Astronomy, University of California, 
Irvine, California 92697 }
\date{\today}
\maketitle

\begin{abstract}

A lithium dopant in a cuprate spin ladder
acts as a vacant (non-magnetic) site accompanied by an
extra hole bound by a Coulomb potential.
We find that, although the undoped ladder 
spin gap is not essentially altered by Li doping, a
dopant-magnon bound state appears within the gap.
Contrary to the case of Zn-doped ladders, we predict that, 
in the Li-doped ladder, the spin liquid character
is very robust against antiferromagnetism.
We also predict the spatial dependence of the density of states in the vicinity of
the dopant which could be measured by local spectroscopic probes.

\smallskip

\end{abstract}

\vskip2pc]

Impurity doping of the high-Tc cuprate superconductors is an
effective tool for exploring the
low temperature physics of these strongly correlated
electronic systems. The similarities or differences observed 
upon doping non-magnetic zinc (Zn) and lithium (Li) ions must 
find their explanations 
both in the strongly correlated nature of their host
and in the peculiarities of each dopant. 
For example, when doped into the antiferromagnetic (AF) phase of La$_2$CuO$_4$,
Li~\cite{AF_Li.LaCuO} 
is far more effective at suppressing the N\'eel temperature
than Zn~\cite{AF_Zn.LaCuO}, although both enter the same planar Cu(2) site.
However, Li, by releasing an 
extra hole (due to the difference of formal valence of
Cu$^{2+}$ and Li$^+$) can theoretically lead to an essential 
perturbation~\cite{skyrmion} while Zn was shown to 
only slightly enhance the AF correlations in its 
vicinity~\cite{AF_Zn.theory}.

Strikingly, despite different behavior in AF systems,
non-magnetic Zn and Li ions behave
quite similarly in conducting YBa$_2$Cu$_3$O$_{6+x}$ (YBCO)
by inducing local magnetic moments~\cite{HiTc_Zn&Li.YBCO}.
This suggests that, in the case of Li, the extra hole is not bound,
unlike La$_2$Cu$_{1-x}$Li$_x$O$_4$.
The induced moment sits predominantly on the four nearest neighbor (NN)
Cu of the dopant and exhibits static~\cite{HiTc_Li.static} 
and dynamic~\cite{HiTc_Li.dynamic} susceptibilities reminiscent 
of a Kondo-like behavior with a large range of Kondo temperatures.
In fact, screening of the induced moment by a conduction
hole resulting in an impurity-hole bound state was 
predicted theoretically~\cite{HiTc_Zn.screening} using a description
in terms of a vacant site embedded in a correlated t-J host.
In superconducting Li-doped YBCO the
moment was found to be weakly modified below Tc~\cite{HiTc_Li.supra}.
Possible reduction of screening due to pairing was also investigated
theoretically~\cite{HiTc_Zn.pair}.

Spin ladders~\cite{ladder_gen} offer a perfect system to 
investigate impurity doping in a spin liquid or Resonating 
Valence Bond~\cite{RVB} (RVB) state,
i.e. a state with short AF correlations and a spin gap. 
Clearly, a reliable microscopic description of dopants in ladders
would greatly contribute to the understanding of their two dimensional (2D) 
analogs as well. In fact, Zn doping into the spin-1/2 Heisenberg 
two-leg ladder SrCu$_2$O$_3$ leads to local moments (as in metallic
YBCO) and to an instability of the gapped spin-liquid state towards 
an AF ordered state~\cite{ladder_exp.Zn} at low enough temperature.
Formation of local moments as well as a rapid suppression of 
the spin gap was obtained theoretically in a Heisenberg ladder 
using the vacant-site impurity model~\cite{ladder_imp.martins}.
Further numerical simulations~\cite{ladder_imp.schollwock}
have led to an
effective model of local moments interacting via an
exchange interaction which rapidly decays with the impurity distance. 

\begin{figure}[htb]
\centerline{\psfig{figure=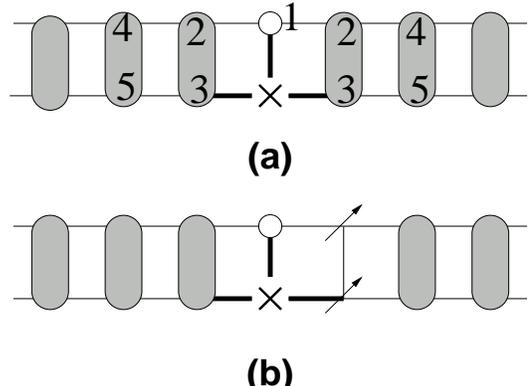,width=0.8\linewidth,angle=0}}
\caption{Schematic representations of a Li-doped spin ladder. The cross
(resp. the circle) stands for the impurity site (resp. injected hole)
and the thick lines corresponds to the attractive potential.
(a) Singlet GS where spins are paired up into spin singlets (shown as
shaded areas). Sites are labelled for convenience.
(b) Lowest triplet excitation: dopant-magnon bound state.
\label{HamiltonFigure}
}
\end{figure}

Doping Li into a two-leg spin ladder is still an open and fascinating 
problem both experimentally and theoretically. 
Novel physics can be expected due to the 
additional (w.r.t the case of Zn) intrinsic doping 
associated with the substitution of a divalent (Cu$^{2+}$) by
a monovalent (Li$^+$) ion. In this Letter, we address this 
issue using a specific vacant-site model for Li$^+$ embedded into
a correlated host. We show that Li doping does not
introduce low-energy spin excitations as in the case of Zn but instead
leads to the formation of a magnon-impurity bound state (BS) just below the 
spin gap of the undoped ladder. This suggests that, unlike Zn-doped ladders, Li-doped
ladders would keep a robust spin liquid character at low
temperature.  

For dilute concentrations, a
single dopant in the geometry of Figs.~\ref{HamiltonFigure} suffices.
We model the substitution of a Cu$^{2+}$ ion by a Li$^+$ atom 
by a vacant inert site which is a hard-core potential for holes.
Because of the different valences of Cu$^{2+}$ and Li$^+$
a hole is injected into the spin ladder and it feels a static attractive
Coulomb potential centered on the Li$^+$ site. 
A realistic Hamiltonian reads:
\begin{eqnarray}
\label{LithiumHamiltonian}
H=J \sum_{<i,j>}^\prime(\mbox{\bf{S}}_{i}\cdot \mbox{\bf{S}}_{j}
&-& \frac{1}{4} \mbox{n}_{i} \mbox{n}_{j} ) \\
 -t \sum_{<i,j>,\sigma}^\prime 
(c_{i,\sigma}^{\dagger}c_{j,\sigma} + \mbox{h.c.})
&-&V \sum_{l_I} (1-\mbox{n}_{l_I}) \, ,\nonumber
\end{eqnarray}
where the notations are standard and the prime means that the sum over 
NN links $<ij>$ is restricted to the bonds
{\it not} connected to the dopant site.
We have modeled the attractive hole potential by a static NN potential
$-V$ as depicted on Figs.~\ref{HamiltonFigure}.
The summation over $l_I$ only runs over the NN sites
of the impurity.
Note that, for simplicity, we restrict ourselves to the case of a magnetically
isotropic ladder i.e. with equal rung and leg couplings, $J$.
Numerical results discussed in the following are obtained
by Exact Diagonalisations (ED) of small {\it periodic} ladder rings 
(up to $2\times 12$) supplemented by Density Matrix Renormalisation 
Group (DMRG) calculations on larger {\it open} systems (up to $2\times 64$).
In open ladders, the impurity site is placed in the center.

\begin{figure}[htb]
\centerline{\psfig{figure=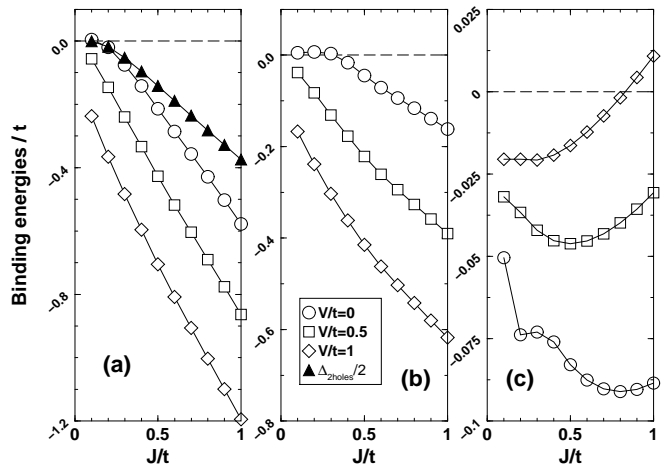,width=1.0\linewidth,angle=0}}
\caption{
Various binding energies to the impurity site 
(see text for definitions) vs $J/t$ obtained by ED on a 2$\times$11 ladder.
The amplitudes  $V$ used for the attractive
potential are indicated on the plot. 
(a) Binding energy of the single hole in the singlet GS. 
For comparison, half of the hole-pair binding energy is also shown.
(b) Binding energy of a hole to the
impurity-magnon system (see~\protect\cite{note_BS2})
in the lowest triplet excitation.
(c) Binding energy of a magnon to the
impurity-hole system (see~\protect\cite{note_BS2}) 
in the lowest triplet excitation.
\label{OneHoleBinding}
}
\end{figure}

First, we investigate the
localization of the injected hole vs the strength 
of the Coulomb potential.
Following Ref.~\cite{HiTc_Zn.screening}, we calculate the so-called
hole-dopant binding energy defined as, 
\begin{eqnarray}
\Delta_{\mbox{\tiny 1$\!$ imp,$\!$ 1$\!$ h}}^{S=0}&=&E_0(1h,1i)+E_0(0h,0i)\\
& & -(E_0(1h,0i)+E_0(0h,1i))\, ,\nonumber
\end{eqnarray}
where $E_0(n\, h,m\, i)$ is the GS energy of a ladder with
$n=0$ or $1$ ($m=0$ or $1$) holes (dopants).
This quantity is negative when the hole and Li ion 
form a stable bound state. 
Since Li-doping results in the simultaneous removal of 
two spins we expect a magnetically inert state
i.e. a singlet ($S=0$). 
As seen on Fig.~\ref{OneHoleBinding}(a) a stable BS 
is found for almost all couplings, even in the case
$V=0$. It should be noticed that the magnitude of 
the binding energy is slightly larger than in the case of a 
2D planar geometry~\cite{HiTc_Zn.screening}.
Fig.~\ref{HamiltonFigure}(a) offers a pictorial representation 
of such a state where the absence of any local moment can be clearly 
understood from the RVB nature of the host (all remaining spins 
are paired up in singlets). 
Although, our calculation shows that a single Li-dopant
binds the injected hole in its vicinity, caution 
is required at a finite concentration.
In this case the possibility of other ``decay channels" has to
be considered, e.g. 2 holes from 2 dopants 
recombining into an itinerant hole pair~\cite{note_BS}.
This possibility is ruled out by the fact that the
dopant-hole binding energy is 
always smaller than half of the hole-pair binding energy 
$\Delta_{\mbox{\tiny 2$\!$ holes}}$ calculated on an undoped
ladder as seen on Fig.~\ref{OneHoleBinding}(a), 
even when $V=0$. 
At low concentrations we then expect decoupled
quenched dopants each binding one hole.
Since the spatial extensions of the isolated BS's 
are quite small (typically $\xi=$ 2 to 4 rungs even when $V=0$), 
the system is expected to remain  insulating up to large doping.

\begin{figure}[htb]
\centerline{\psfig{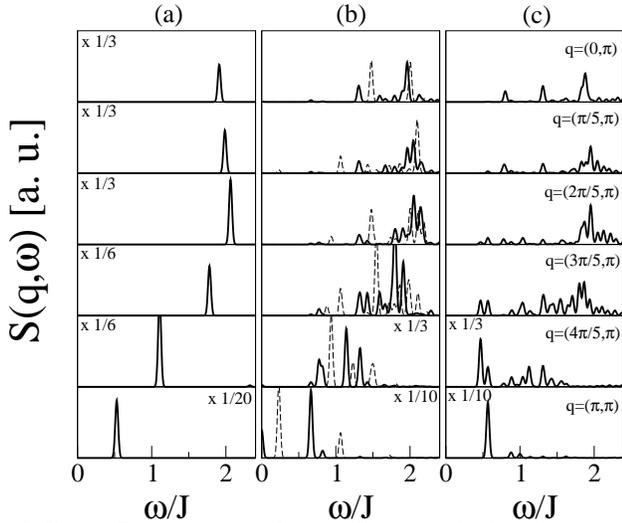}}
\caption{
Spin structure factors $S({\bf q},\omega)$ 
calculated on $2\times 10$ ladders. Data for $q_\perp=\pi$ only are shown.
From bottom to top, the different curves correspond to decreasing 
$q_\parallel$ momenta, from ${\bf q}=(\pi,\pi)$ (bottom) to 
${\bf q}=(0,\pi)$ (top).
For clarity, reducing scaling factors (as indicated) are applied on
some curves. 
(a) Undoped periodic ladder; (b) Spin ladder doped with a single Zn 
impurity (full line) or two Zn impurities 
separated by the maximum distance on the same leg (dashed line); 
(c) Ladder doped with a single Li impurity with $t=2J$ and $V/t=0.5$.
\label{Magn_factor}
}
\end{figure}

We now turn to the magnetic properties of the Li-doped spin ladder
in order to compare with Zn doping. 
For this purpose, we compute the dynamical spin structure
factor $S({\bf q},\omega)$ as measured in Inelastic Neutron Scattering (INS) experiments
for an undoped ladder (shown on Fig.~\ref{Magn_factor}(a)), a Zn-doped ladder
(modeling Zn as a neutral vacant site and shown on Fig.~\ref{Magn_factor}(b))
and a Li-doped ladder described by Eq.~(\ref{LithiumHamiltonian}) 
(shown on Fig.~\ref{Magn_factor}(c)). 
Note that $S({\bf q},\omega)$ quantifies the
spectral weight given by the matrix element 
$|\big<n|S_{\bf q}^Z|0\big>|^2$ between the singlet
GS $|0\big>$ and the magnetic excitation $|n\big>$ at energy $\omega$.
The RVB picture, in which spins are paired up
into short range singlets (for example on the
rungs as in Figs.~\ref{HamiltonFigure}), turns out to give a nice
qualitative understanding of the magnetic properties. 
In the undoped ladder, a triplet excitation (magnon)
is well described by exciting a rung singlet into a triplet.
Fig.~\ref{Magn_factor}(a) clearly shows the single magnon dispersion
with a minimum at ${\bf q}=(\pi,\pi)$ which defines the spin gap
$\Delta_S^{0}$~\cite{sg} of the undoped ladder. Introducing a Zn atom 
on a rung releases a free spin-1/2 which leads to zero-energy spin 
fluctuations predominantly at ${\bf q}=(\pi,\pi)$ as seen in 
Fig.~\ref{Magn_factor}(b) while the undoped ladder magnon 
mode still survives.
When two Zn dopants are introduced, our data, in agreement 
with Ref.~\cite{ladder_imp.martins}, show that the two local S=1/2 moments 
experience a weak effective exchange interaction, $J_{\mbox{\small eff}}$, 
which decays rapidly with separation. 
In that case, a small spectral weight at ${\bf q}=(\pi,\pi)$ and low 
energy $\omega\sim J_{\mbox{\small eff}}$ below the undoped spin gap is seen.
The case of Li-doping shown on Fig.~\ref{Magn_factor}(c) is drastically
different with no weight at small energy.
Indeed, since a Li$^+$ dopant introduces
simultaneously a {\it bound} hole, it does not release a free
spin as for Zn but rather produces a new type of excitation
located just below the unperturbed spin gap, namely a 
collective bound state of a (let us say rung) magnon with the hole and 
the Li ion as naively depicted on Fig.~\ref{HamiltonFigure}(b).
Its binding energy defined as the energy separation w.r.t. the
free magnon energy $\Delta_S^{0}$ remains in general quite small 
(in absolute value) as seen 
directly in Fig.~\ref{Magn_factor}(b) (and qualitatively in 
Fig.~\ref{OneHoleBinding}(c)). Therefore a drastic 
reduction of the spin gap does not occur in this case.

\begin{figure}[htb]
\centerline{\psfig{figure=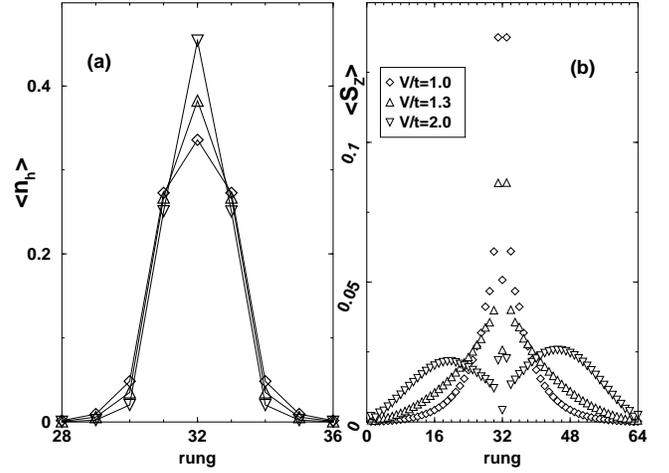,width=1.0\linewidth,angle=270}}
\caption{
Hole rung density (a) and $S_z$ rung density (b) along the ladder direction 
in the lowest energy triplet state
calculated by DMRG for $J/t=0.5$. The rung density is defined as the algebraic 
sum of the densities (if any) on the two sites of a given rung. Different values 
of $V/t$ (as indicated) are shown. 
\label{MagBound}
}
\end{figure}

The physical origin of this BS is of particular interest. 
Simple qualitative arguments as well as extensive numerical data
show that the origin of binding in this channel can be attributed
to the hole kinetic energy gain associated with the spin polarization.
DMRG calculations show that the spatial extents of the charge and 
spin perturbations associated with the BS are quite different.
While the hole is localized on a scale of a few rungs from
the dopant as seen in Fig.~\ref{MagBound}(a), the distribution of 
the rung magnetization can extend to large distances
as seen in Fig.~\ref{MagBound}(b). 
This is in agreement with our other finding 
that the energy costs (i.e. the absolute value of the
related binding energies~\cite{note_BS2}) of the two virtual
decay processes leading either 
to a free hole and a neutral dopant-magnon complex 
or to a free magnon and a dopant-hole complex 
are quite different
in magnitude as seen in Figs.~\ref{OneHoleBinding}(b) and
\ref{OneHoleBinding}(c).
In fact, the attraction $-V$ has the opposite effect on the two quantities;
while increasing $V$ binds the hole more strongly, it also limits the ability
of the hole to reduce its kinetic energy by moving in the spin
polarized background of the magnon, and so the
binding energy of the magnon to the dopant-hole complex is reduced~\cite{note_tp}. 
From the analysis of the behavior of the magnetic size of the BS 
(deduced from Fig.~\ref{MagBound}(b)) or alternatively
directly from its binding energy plotted on
Fig.~\ref{OneHoleBinding}(c), we conclude that above a
critical value of $V$ (typically one gets $V_C/t\simeq 2$ for $J/t=0.5$) 
the magnon escapes from the dopant. 

\begin{figure}[htb]
\centerline{\psfig{figure=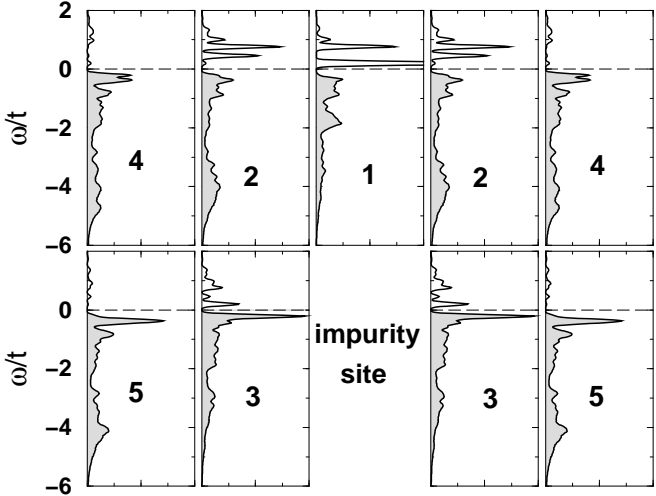,width=1.0\linewidth,angle=0}}
\caption{
Local DOS around a Li dopant obtained by ED of a
$2\times 9$ ladder with $J/t=0.5$ and $V/t=0.5$. Each panel corresponds to
a site in the vicinity of the dopant (site labels correspond
to those of Fig.~\protect\ref{HamiltonFigure}(a)). 
Occupied (empty) electronic states are shaded (left blank). 
Note that, for a dopant concentration $x$, the total $\omega>0$ (resp. 
$\omega<0$) integrated weight scales as $x$ (resp. $\frac{1}{2}-\frac{x}{2}$).
\label{DOS_STS}
}
\end{figure}

For completeness, we calculate also the local density of state (DOS)
in the vicinity of the dopant. Since this DOS might be 
accessible experimentally by Scanning Tunneling Spectroscopy
it can test the validity of the model.
Results are shown in Fig.~\ref{DOS_STS} corresponding to 
spatially resolved DOS spectra in the vicinity of the dopant. 
The $\omega>0$ ($\omega<0$) spectra give the weights 
of the neutral (charged) target $S=1/2$ states which are accessible 
by removing the hole (adding an extra hole) in the singlet GS. 
The large peak seen at vanishing positive energy on site number 1 
i.e. on the same rung as the dopant corresponds to 
a resonant magnetic spinon-dopant BS. Other spinon-dopant
resonances of higher energies are also seen with larger 
spatial ranges i.e. spread over two or three sites of the leg opposite to
the dopant. Similarly, $\omega<0$ resonances are seen
when adding an extra hole {\it on the same leg} as the dopant 
e.g on sites number 3 and 5.
The lowest resonance energy (in absolute value) is obtained
when the second hole is added on sites number 3 next to the dopant.
Note also that the local $\omega>0$ ($\omega<0$) integrated weight 
provides directly the local hole (electron) density in the GS.
Hence our data show that the bound hole is located mainly on the 
opposite leg to the dopant and extends roughly over three sites.

Our theory can be directly tested, if Li can be substituted for Cu in  the 
ladder compound, SrCu$_2$O$_3$. The extra bound hole around a Li dopant should 
ensure that a free local S=1/2 moment is not created in contrast 
to the case of Zn 
doping. Hence we predict that SrCu$_{2-x}$Li$_x$O$_3$ should not order 
antiferromagnetically at low temperature unlike SrCu$_{2-x}$Zn$_x$O$_3$
(Ref.~\onlinecite{ladder_exp.Zn}). Further the 
nature of the magnon-dopant BS, the charge distribution and local DOS could 
also be examined experimentally in these systems. However, our analysis 
raises interesting questions regarding the close similarity between Li 
and Zn substitution reported in superconducting YBCO samples. In particular 
if we interpret the spin gap phase in underdoped YBa$_2$Cu$_3$O$_{6.6}$ as a doped 
d-RVB phase, then there should be a close similarity to the behavior of the 
doped ladder. However Bobroff et al.~\cite{HiTc_Li.static,HiTc_Zn&Li.YBCO} 
report a free S=1/2 appears for both Zn and Li doping of these samples. 
A possible way to reconcile this apparent contradiction is to postulate that
Li$^+$ does not bind a hole in YBa$_2$Cu$_{3-x}$Li$_x$O$_{6.6}$, 
but does in La$_2$Cu$_{1-x}$Li$_x$O$_4$. This could occur if the mobile 
O$^{2-}$-ions in the chains were repelled from the Li$^+$-ions on the planes. A test of 
this hypothesis can be made by doping Li$^+$ and Zn$^{2+}$ in YBa$_2$Cu$_4$O$_8$ 
which as a stoichiometric compound has no mobile O-ions. Our analysis then 
predicts free S=1/2 moments only for Zn-doping and not for Li-doping in this case.

In conclusion, our analysis predicts a clear distinction between the magnetic 
properties of the two non-magnetic ions, Zn$^{2+}$ and Li$^+$ when doped into 
spin liquids due to the binding of a hole in the latter case. 
Experiments to test these predictions are proposed. 

We thank Matthias Troyer for valuable discussions. SRW thanks
the NSF for support under grant DMR98-70930. DP also acknowledges support
from the Center for Theoretical Studies at ETH-Z\"urich.


\begin{references}
\bibitem{note_BS} Note that the other decay channel consisting of 
a single dopant trapping two holes can also be rejected on 
energetic grounds since the two hole-dopant
binding energy $\Delta_{\mbox{\tiny 1 imp., 2 holes}}^{S=0}$
defined as $E(2h,1i)+E(0h,1i)-2E(1h,1i)$ was always found to be
positive.
\bibitem{note_BS2} The binding energies related to these two decay
processes and plotted in Figs.~\protect\ref{OneHoleBinding}(b),(c) can be 
defined as 
$\Delta_{\mbox{\tiny 1$\!$ imp,$\!$ 1$\!$ h}}^{S=1}
=E_1(1h,1i)+E_0(0h,0i)-(E_0(1h,0i)+E_0(0h,1i))$ 
and as 
$\Delta_S-\Delta_S^{0}=
E_1(1h,1i)+E_0(0h,0i)-(E_0(1h,1i)+E_1(0h,0i))$ 
where $E_1$ is the energy in the triplet sector.
\end{references}
\end{document}